\documentclass[aps,prl,twocolumn,showpacs,groupedaddress]{revtex4}
\usepackage{graphicx}
\begin{document}

\title{Evidence for structural crossover in the supercritical state}
\author{Dima Bolmatov$^{1}$}
\thanks{d.bolmatov@gmail.com, db663@cornell.edu}
\author{V. V. Brazhkin$^{3,4}$}
\author{Yu. D. Fomin$^{3}$}
\author{V. N. Ryzhov$^{3,4}$}
\author{K. Trachenko$^{2,5}$}
\address{$^1$ Baker Laboratory, Cornell University, Ithaca, New York 14853-1301, USA}
\address{$^2$ School of Physics and Astronomy, Queen Mary University of London, Mile End Road, London, E1 4NS, UK}
\address{$^3$ Institute for High Pressure Physics, RAS, 142190, Troitsk, Moscow Region, Russia}
\address{$^4$ Moscow Institute of Physics and Technology, 141700 Moscow, Russia}
\address{$^5$ South East Physics Network}

\begin{abstract}
The state of matter above the critical point is {\it terra incognita}, and is loosely discussed as a physically homogeneous flowing state where no differences can be made between a liquid and a gas and where properties undergo no marked or distinct changes with pressure and temperature. In particular, the structure of supercritical state is currently viewed to be the same everywhere on the phase diagram, and to change only gradually and in a featureless way while moving along any temperature and pressure path above the critical point. Here, we demonstrate that this is not the case, but that there is a well-defined structural crossover instead. Evidenced by the qualitative changes of distribution functions of interatomic distances and angles, the crossover demarcates liquid-like and gas-like configurations and the presence of medium-range structural correlations. Importantly, the discovered structural crossover is closely related to both dynamic and thermodynamic crossovers operating in the supercritical state, providing new unexpected fundamental interlinks between the supercritical structure, dynamics and thermodynamics.
\end{abstract}
\pacs{05.70.Fh, 65.20.Jk, 61.25.Mv}

\maketitle
It is currently believed that the supercritical state is physically homogeneous, in the sense that moving along any path on a pressure and temperature phase diagram above the critical point does not involve any marked or distinct changes of properties which vary only gradually and in a featureless way \cite{hansen,su1,su2}. The recent explosion of the new ways and applications where supercritical fluids are deployed has called for better fundamental understanding of the supercritical state of matter \cite{su1,su2}. In particular, further deployment is seen to be limited by the absence of solid theoretical guidance \cite{su1,su2}.

The structure of supercritical state is currently viewed to be the same everywhere on the phase diagram and to change only gradually and in a featureless way while moving along any path on a pressure and temperature phase diagram above the critical point \cite{su1,su2}. Here, we demonstrate that this is not the case, but that there is a well-defined structural crossover instead. Our detailed analysis reveals the crossover between the structure with well-defined short- and medium-range order as in liquids and the random gas-like structure. We explain the origin of this crossover, and relate its origin to the change of dynamic and thermodynamics properties of the supercritical state. Our analysis reveals new interesting features that bear important consequences for the ongoing effort in elucidating and understanding the structure and properties of disordered matter such as liquids and glasses \cite{billinge,dyre,ryl,harow,biroli,err,wer,salmon,chandler,ma,jstat,jap,widom,moor,wei,symm}.

\begin{figure}
	\centering
\includegraphics[scale=0.6]{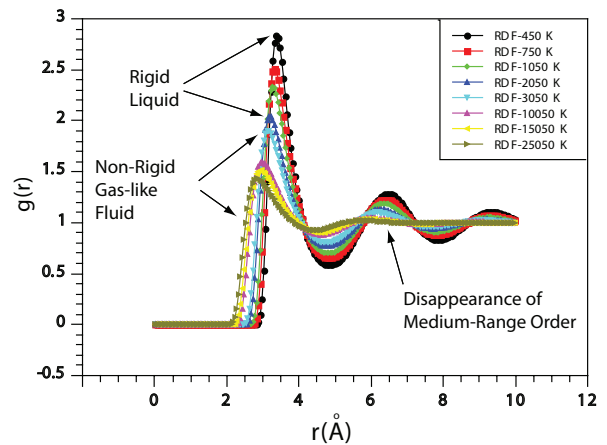}
\caption{The radial distribution functions $g(r)$ of simulated one-component Lennard-Jones (LJ) supercritical fluid at different temperatures showing the disappearance of the medium-range order at high temperature. The simulations are performed at constant density.}
	\label{fig1}
\end{figure}

Using Molecular Dynamics (MD) simulations \cite{todorov1}, we have simulated one-component Lennard-Jones (LJ) fluid fitted to Ar properties \cite{martin}. We have simulated the system with 32000 atoms using constant-volume (NVE) ensemble in the very wide temperature range (see Figs. 1--3) well extending into the supercritical region; the system was equilibrated at constant temperature. The temperature range in Figs. 1-3 is between about 3$T_c$ and 167$T_c$, where $T_c$  is the critical temperature of Ar, $T_{c}\simeq$150 K (1.3 in LJ units). The simulated density, 1880 kg/m$^3$ (1.05 in LJ units), corresponds to approximately three times the critical density of Ar. A typical MD simulation lasted about 50 ps, and the properties were averaged over the last 20 ps of simulation, preceded by 30 ps of equilibration. The simulations at different temperature included 500 temperature points simulated on the high-throughput computing cluster. We calculate radial distribution function (RDF), average it over the last 20 ps of the simulation, and show its temperature evolution in Figure 1. We observe the decrease of the first peak of the RDF, and the near disappearance of the second and third peaks of RDFs, implying that the medium-range structure is no longer visible at high temperature.

\begin{figure*}
	\centering
\includegraphics[scale=0.5]{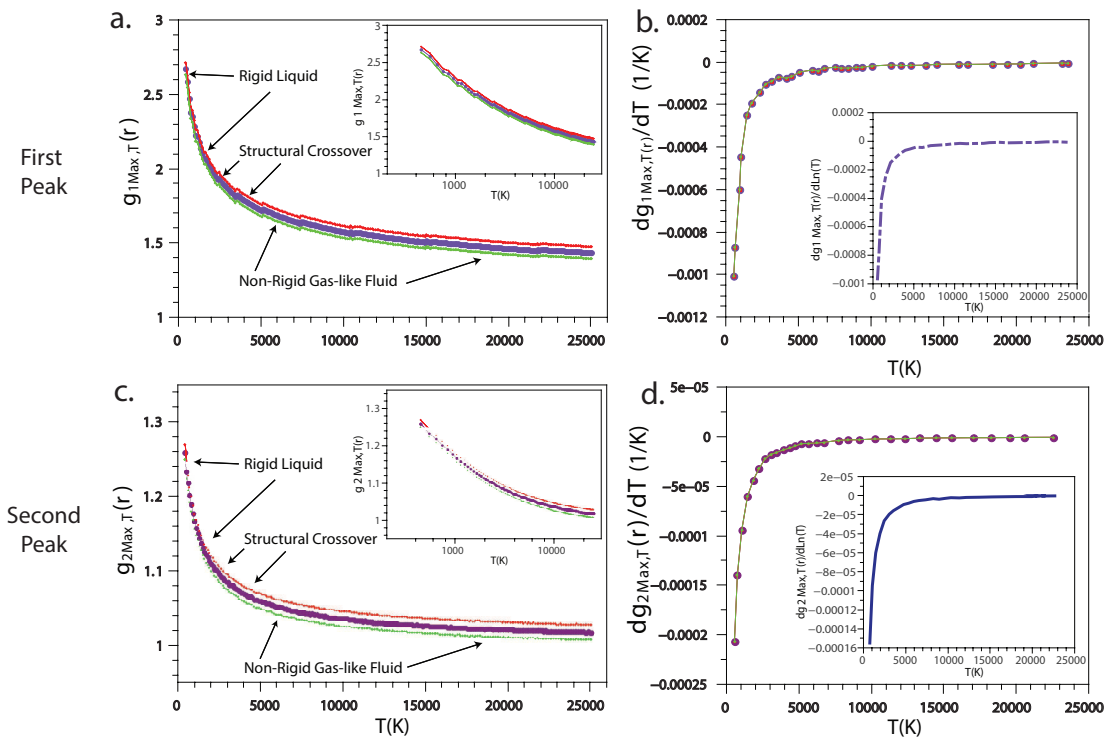}
\caption{Evolution of the first and second peaks of RDF (thick lines) of the supercritical state with temperature (a,c) and their derivatives (b,d). Thin lines show the error in simulations, calculated from fluctuations of RDF at each temperature. The inset shows the same in the logarithmic plot.}
	\label{fig2*}
\end{figure*}

To analyze temperature changes of RDF in more detail, we calculate the heights of the first and second peaks of RDF, and plot these in Figure 2 as a function of temperature. We observe the steep decrease of both peaks at low temperature, followed by their flattening at high temperature, with the crossover between the two regimes taking place around 3000 K. To make the crossover more visible, in Figures 2b and 2d we plot the temperature derivative of the heights of both peaks. These plots clearly show two regimes corresponding to the fast and slow change of RDF peaks.

The structural crossover is further evidenced by the calculation of interatomic angles. For each atom $i$, we define its nearest neighbours and calculate all possible angles that each pair of neighbouring forms with the central atom $i$. The distribution of these angles has several maxima. For definitiveness, we select the angle that corresponds to the first maximum, $\alpha_m$. For a range of different densities, we calculate $\alpha_m$ at different temperatures, and subsequently average it over time and different local configurations. In Figure 3, we show the temperature dependence of averaged $\alpha_m$ in re-scaled units of both density and temperature in order to compare the simulations at different density in one plot. Figure 3 shows the same trend as the heights of the RDF peaks above: the steep initial decrease of $\alpha_m$ from about 60 degrees (the characteristic angle in the face-centred cubic lattice), followed by flattening and constancy at high temperature.

We have repeated the same simulations in the constant pressure ensemble at various pressures, and have found the same crossover as the one shown in Figures 1--2, except the crossover temperature decreases. For example, at 3 GPa, the crossover operates in the temperature range of about 1500--2500 K. This is an important insight with regard of future experimental detection of the structural crossover in the supercritical state.

The structural crossover in the supercritical state is the new effect not hitherto anticipated, in view of the currently perceived physical homogeneity of the supercritical state \cite{hansen,su1,su2}. We now address the origin of this crossover, and relate this origin to the changes of both dynamics and thermodynamics of the supercritical state.

\begin{figure}
	\centering
\includegraphics[scale=0.3]{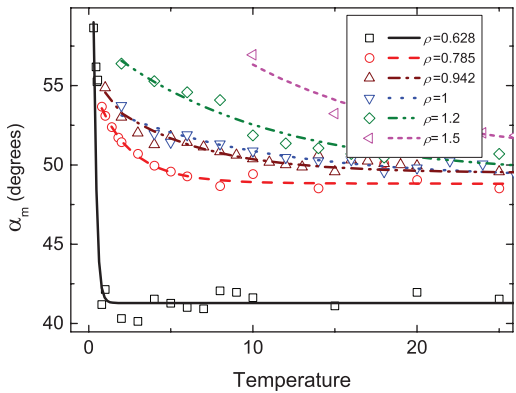}
\caption{Evolution of $\alpha_m$ with temperature for different densities in the supercritical state. Density and temperature are in LJ units where $\rho=1$ corresponds to approximately 1880 kg/m$^3$ and $T=1$ to 115 K for Ar.}
	\label{angle}
\end{figure}

We have recently proposed that a new line exists on the phase diagram {\it above} the critical point extending to arbitrarily high pressure and temperature \cite{pre,j2,natcomm,prl}. This line, the Frenkel line, separates two dynamically distinct states.  Particles in a gas move in almost straight lines until they collide with other particles or container walls and change course. In liquids, particle motion has two components: a solid-like,  quasi-harmonic vibrational motion about equilibrium locations and diffusive jumps between neighboring equilibrium positions \cite{frenkel}. As the temperature increases or the pressure decreases, a particle spends less time vibrating and more time diffusing. Eventually, the solid-like oscillating component of motion disappears; all that remains is the ballistic-collisional motion. That disappearance, a qualitative change in particle dynamics, corresponds to the dynamic crossover at the Frenkel line. We have shown that crossing the Frenkel line corresponds to the qualitative change of atomic structure in a liquid, the transition of the substance from the ``rigid'' liquid structure to the ``non-rigid'' gas-like structure (see Figures 1-3 and discussion below). This transition takes place when liquid relaxation time $\tau$ ($\tau$ is liquid relaxation time, the average time between two consecutive atomic jumps at one point in space \cite{frenkel}) approaches its minimal value, $\tau_{\rm D}$ , the Debye vibration period. Crossing the Frenkel line is accompanied by qualitative changes of most important properties of system, including diffusion, viscosity, thermal conductivity and dispersion curves \cite{pre}. Recently, we have
discovered that specific heat shows a crossover between two different regimes \cite{natcomm}. We subsequently formulated a theory of system thermodynamics above the crossover, derived a power law and analyzed supercritical scaling exponents in the system above the Frenkel line. In this theory, energy and heat capacity are governed by the minimal length of the longitudinal mode in the system only, and do not explicitly depend on system-specific structure and interactions \cite{natcomm}. More recently, we have proposed a mathematically rigorous way to define the Frenkel line as the point on the phase diagram where the oscillations of the velocity correlation function disappear \cite{prl}.

An important change, to which we return below, is the ability of the system to support high-frequency shear modes with $\omega>\frac{1}{\tau}$ (predicted by Frenkel \cite{frenkel}, this ability was later confirmed experimentally \cite{pilgrim,monaco,hosoka}): when $\tau\approx\tau_{\rm D}$, the system loses the ability to support shear modes at all available frequences, and behaves like a gas. We therefore call the system as "rigid" liquid below the Frenkel line, and "non-rigid" gas-like fluid above the line (see Figures 1--4).

\begin{figure*}
	\centering
\includegraphics[scale=0.5]{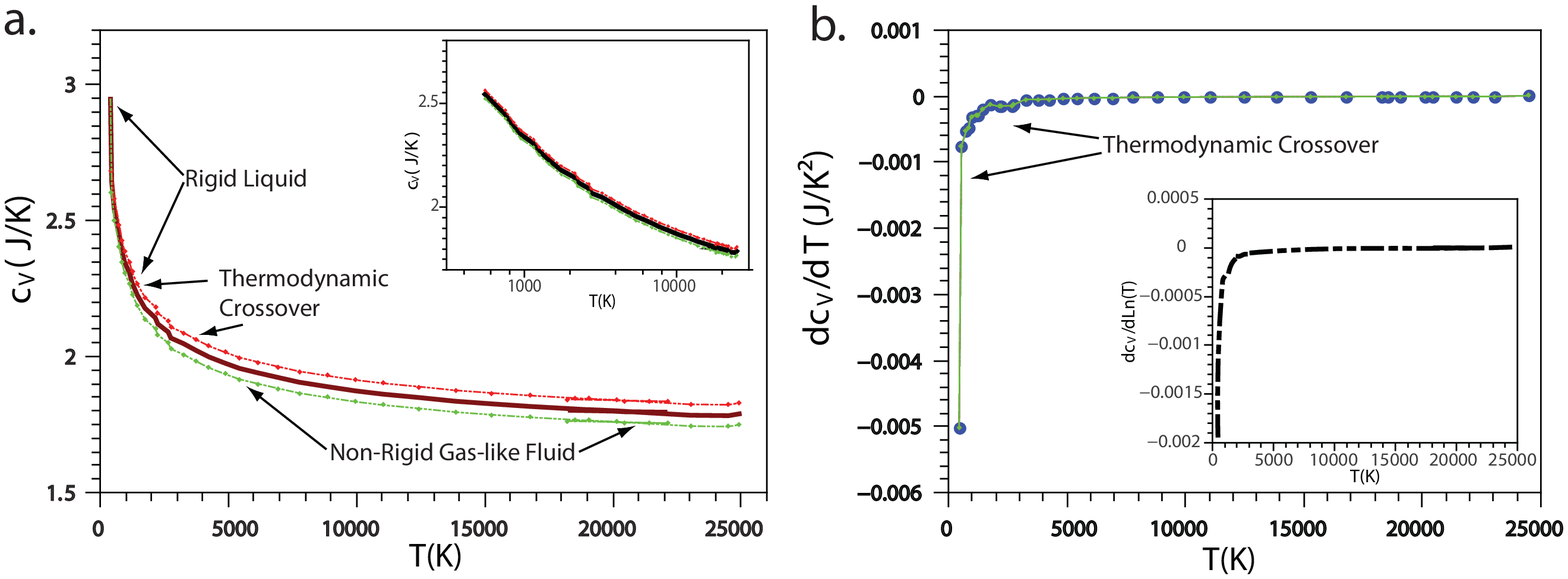}
\caption{ Calculated $c_{V}$ (thick lines) showing the crossover and continuous thermodynamic transition around $c_{V}\simeq$ 2, ($k_{\rm B}=$ 1). The crossover takes place between different dynamic regimes of the rigid liquid and non-rigid supercritical fluid. The thermodynamic crossover undergoes within 2000~K-4000~K temperature range granting the relationship between thermodynamic, dynamic and the structural crossover (see Figs. 1-3) correspondingly. Thin lines show the error in simulations, calculated from fluctuations of energy at each temperature. The inset shows the same in the logarithmic plot.}
	\label{cv}
\end{figure*}

The structural crossover seen in Figures 2--3 can now be explained as follows. In the regime where $\tau\gg\tau_{\rm D}$, particles oscillate for long time before jumping to the neighbouring quasi-equilibrium sites. This means that the short- and medium-range order similar to that present in amorphous systems such as glasses or viscous liquids is maintained during time $\tau$. Consequently, the calculated RDF shows the correlations due to atomic packing in several spheres that extend beyond the short-range order as in subcritical liquids, as is seen in Figure 1 at low temperature. When $\tau$ approaches its minimal value of approximately $\tau_{\rm D}$, the oscillatory component of the motion is lost, and the dynamics becomes ballistic as in a gas. The gas-like configuration has qualitatively different structure with notably less constraints and structural correlations in the medium range, as is seen in Figure 1. As a result, the structural crossover witnessed by the RDF peaks is intimately related to the dynamic crossover at the Frenkel line.

Importantly, RDF peaks on both sides of the crossover are predicted to change differently with temperature in this picture. Indeed, in the rigid liquid regime, RDF peaks decrease rapidly due to the fast exponential decrease of $\tau$ and the accompanied decrease of particles located at similar distances. On the other hand, in the gas-like regime where the oscillatory motion and the medium-range order are no longer present, structural correlations are less sensitive to temperature increase because the dynamics is already randomized by ballistic motions as in a gas. The temperature dependence shown in Figure 2 is consistent with this picture, and supports the relationship between the {\it structure and dynamics} of the supercritical state discussed above.

\begin{table*}[ht]
\centering
\begin{tabular}{c c  c c c c}
\hline\hline
State & Structure  & Dynamics & $<E_{kin}>$ & $<E_{pot}>$ & $C_{V}$\\ [0.5ex]
& & & $k_{\rm_B}NT$ & $k_{\rm_B}NT$ & $k_{\rm_B}N$\\ [0.5ex]
\hline
{\begin{tabular}{ll}Solid\end{tabular}} & Medium and/or  Long-Range Order & 1-L+2-S Phonon Modes & $\frac{3}{2}$ & $\frac{3}{2}$ & $3$ \\
{\begin{tabular}{ll}Liquid\end{tabular}} & Medium and/or  Short-Range Order  & 1-L+2-S ($\omega_{s} >\frac{1}{\tau}$) Phonon Modes & $\frac{3}{2}$ & $\frac{3}{2}$ $\rightarrow$ $\frac{1}{2}$ & $3$ $\rightarrow$ $2$ \\
{\begin{tabular}{ll}Ideal \\ Gas\end{tabular}} & No Order  & No Phonon Modes & $\frac{3}{2}$ & 0 & $\frac{3}{2}$ \\
\hline\hline
{\begin{tabular}{ll}Rigid \\ Liquid\end{tabular}} & Medium-Range Order  & 1-L+2-S ($\omega_{s} >\frac{1}{\tau}$) Phonon Modes & $\frac{3}{2}$ & $\frac{3}{2}$ $\rightarrow$ $\frac{1}{2}$ & $3$ $\rightarrow$ $2$ \\
\hline
{\begin{tabular}{ll}Non-Rigid \\ Gas-like Fluid \end{tabular}} & Short-Range Order   & 1-L Mode & $\frac{3}{2}$ & $\frac{1}{2}$ $\rightarrow$ 0 & $2$ $\rightarrow$ $\frac{3}{2}$  \\
[1ex]
\hline\hline
\end{tabular}
\caption{Structural, dynamic and thermodynamic properties of solids, liquids and gases, followed by two states in the supercritical region: Rigid Liquid and Non-Rigid Gas-like Fluid. L and S letters refer to longitudinal and shear modes.}
\label{table1}
\end{table*}

In addition to the relationship between structural and dynamic crossovers discussed above, there is also an important link between structural and thermodynamic crossovers of the supercritical system. In Figure 4, we show the calculated constant-volume specific heat, $c_v=\frac{1}{N}\left(\frac{\partial E}{\partial T}\right)_v$, of the simulated system as a function of temperature using the same interatomic LJ potential. We observe a crossover of $c_v$ around 3000 K, an effect that is also clearly seen in Figure 4b where the temperature derivative of $c_v$ is shown. Importantly, the crossover of $c_v$ takes place at the same temperature as the crossover of RDF peaks in Figure 2. This provides the quantitative evidence for the close relationship between the {\it structural and thermodynamic} properties and their crossover in the supercritical system, the relationship we explore below.

We first note that the crossover of $c_v$ in Figure 4, $c_v\approx 2k_{\rm B}$, is non-coincidental. We have recently provided the quantitative theory of liquid $c_v$ \cite{prbA,scirep}. In this theory, the reduction of heat capacity from $3k_{\rm B}$ to $2k_{\rm B}$ is due to the progressive loss of two shear modes with frequency $\omega>\frac{1}{\tau}$. When all shear modes are lost at the Frenkel line as discussed above, only the remaining longitudinal mode with the potential energy of $\frac{1}{2}Nk_{\rm B}T$ contributes to the total system energy. Together with $\frac{3}{2}Nk_{\rm B}T$ given by kinetic energy, the total energy is $2Nk_{\rm B}T$, giving $c_v=2k_{\rm B}$ \cite{prbA,scirep}. On further temperature increase when $c_v$ decreases from $c_v=2k_{\rm B}$ to its ideal-gas value of $c_v=3/2k_{\rm B}$, another mechanism kicks in: the disappearance of the remaining longitudinal mode with the wavelength smaller than the mean-free path of particles \cite{natcomm}. The two mechanisms naturally give a crossover of $c_v$ at $c_v=2k_{\rm B}$.

In addition to the evidence from MD simulations, the relationship between the crossover of structure and thermodynamics in the supercritical states can also be discussed on the basis of the general relationship between the system energy, $E$, and $g(r)$:

\begin{equation}
E=\frac{3}{2}Nk_{\rm B}T+2\pi N\rho\int_{0}^{\infty}r^{2}u(r)g(r)dr
\end{equation}

\noindent where $N$ is the number of particles, $\rho$ is the density and $u(r)$ is the interatomic potential.

According to Eq. (1), the crossover of $g(r)$ necessarily implies the crossover of energy and $c_v$ and vice versa, consistent with our observation that the crossovers of RDFs seen in Figure 2 is accompanied by the crossover of $c_v$ in Figure 4. Physically, these crossovers are related as follows. In the rigid-liquid supercritical regime when $\tau\gg\tau_{\rm D}$, the initial steep decrease of $c_v$ from 3$k_{\rm B}$ to 2$k_{\rm B}$ is due to the fast loss of two shear modes as discussed above. This decrease is fast because $\tau$ exponentially decreases with temperature, giving fast decrease of $c_v$ seen in Figure 4. At the same time, $\tau\gg\tau_{\rm D}$ implies that solid-like structures persist during time $\tau$ and hence give rise to the correlations in the medium range as seen in Figure 1, correlations that decrease steeply in Figure 2. In the non-rigid gas-like supercritical regime, $c_v$ decreases slowly due to the slow disappearance of the remaining longitudinal mode in the gas-like state \cite{natcomm}, which via Eq. (1) implies slowly decaying behavior of $g(r)$, consistent with Figure 2. The two effects explain why both $c_v$ and $g(r)$ undergo a crossover simultaneously, as Eq. (1) demands.

We note that the relationship between structure and dynamics have always fascinated scientists in condensed matter physics, with the recognition that such a relationship may exist in some classes of systems but not in others \cite{martin,kitay,gotze}. Here, we find that not only the supercritical state is not physically homogeneous as previously viewed, it is also universally amenable to supporting fundamental interlinks between all three fundamental system properties: structure, dynamics and thermodynamics. This is summarized in Table 1.

In summary, we presented the evidence for the existence of the structural crossover in the supercritical state of matter, and discussed the relationship between the structure, dynamics and thermodynamics of the supercritical state. The existence of this crossover has not been hitherto anticipated, and is contrary to how the supercritical state has been viewed until now. Our results therefore call for the experiments (neutron, X-ray scattering experiments) to detect the crossover.

\section{Acknowledgements} D. Bolmatov thanks Ben Widom and Cornell University for support. We are grateful to A. Lyapin for discussions.

\end{document}